\documentclass{llncs}

\usepackage{url}
\usepackage{amssymb}
\usepackage{amsmath}

\def\avgA{\ensuremath{\mathit{\mu_A}}}
\def\avgS{\ensuremath{\mathit{\mu_S}}}
\def\NMI{\ensuremath{\mathit{NMI}}}
\def\nSW{\ensuremath{\mathit{n_{SW}}}}
\def\nW{\ensuremath{\mathit{n_{W}}}}
\def\nA{\ensuremath{\mathit{n_{A}}}}

\begin{document}
\title{Clustering students' open-ended questionnaire answers}
\titlerunning{Clustering questionnaire answers}  

\author{Wilhelmiina H\"am\"al\"ainen{1} \and Mike Joy{2} \and Florian Berger{3} \and Sami Huttunen{4}}
\authorrunning{H\"am\"al\"ainen et al.}
\tocauthor{Wilhelmiina H\"am\"al\"ainen, Mike Joy, Florian Berger, Sami Huttunen}
\institute{Aalto University, Department of Computer Science, P.O. Box 15400, FI-00076 Aalto, Finland,\\
\email{whamalai@iki.fi}
\and
University of Warwick, Department of Computer Science, CV4 7AL Coventry, United Kingdom\\
\email{m.s.joy@warwick.ac.uk}
\and 
Vincit, Visiokatu 1, FI-33720 Tampere, Finland\\
\email{florian.berger@vincit.fi}
\and 
SKJ Systems Ltd Oy, Sammonkatu 12, FI-70500 Kuopio, Finland\\
\email{sami.huttunen@skj.fi}}

%
%
%

\maketitle 

\begin{abstract}
Open responses form a rich but underused source of information in
educational data mining and intelligent tutoring systems. One of the
major obstacles is the difficulty of clustering short texts
automatically. In this paper, we investigate the problem of clustering
free-formed questionnaire answers. We present comparative
experiments on clustering ten sets of open responses from course
feedback queries in English and Finnish. We also evaluate how well the
main topics could be extracted from clusterings with the HITS
algorithm. The main result is that, for English data, affinity
propagation performed well despite frequent outliers and considerable
overlapping between real clusters. However, for Finnish data, the
performance was poorer and none of the methods clearly outperformed
the others. Similarly, topic extraction was very successful for the
English data but only satisfactory for the Finnish data. The most
interesting discovery was that stemming could actually deteriorate the
clustering quality significantly.
\end{abstract}
\section{Introduction}

Receiving feedback from students is an essential part of the modern
educational process, but dealing with the responses from large classes
can be time-consuming. Open-ended questions in feedback forms often
provide the most detailed and accurate information, but analyzing
students' answers is potentially a laborious task which may require the
application of non-trivial qualitative data analysis techniques.

With open response questions, students are not forced to approximate
their real answers with pre-fixed choices, and they can also reveal
extra information either explicitly (by answering further questions)
or implicitly (by use of word-choices and syntactic structures). These
kinds of answers are especially desirable when gathering qualitative
information, such as on students' motivation and attitudes. However,
analyzing open responses is laborious for a human evaluator and very
challenging with existing data mining and data analysis tools. It
therefore comes as no surprise that in educational data mining the
standard solutions have been (a) to omit open response variables from
the analysis, and (b) to use only closed questions with pre-fixed
answer categories (even for querying attitudes).

In this paper, we investigate better solutions for analyzing open response
questions automatically, in order to speed up and improve the processing of
student feedback data. In particular, we are interested in how to
cluster short, free-formed textual questionnaire answers. Loosely
speaking, clustering means dividing data points into a set of groups,
such that points in each group are similar or close to each other but
different or distant from points in the other groups. This is exactly
what a human analyzer would be likely to do with such data: they would
divide the answers into categories to see a summary of how students are
doing, what their main messages are, and whether there are individuals
or subgroups who would require extra attention. All this information can
be used for modifying a course and targeting learning and teaching issues.

In previous research, there have been many instances of how clustering
of student responses (by variables other than text) can be utilized in
education. Most of them have used the clustering information for
descriptive purposes (understanding the data), such as identifying
successful learning patterns \cite{kaser} or effective ways of using
learning tools \cite{perera}, allocating students into different
teaching groups \cite{huikkola} or targeting tutoring
\cite{schmitt2007}. On the other hand, clustering can also be an
important step in the construction of predictive models for
intelligent tutoring systems. For example, clustering can help to
identify natural classes and features which separate those classes
effectively 
in the construction of a $K$-nearest neighbor style of classifier
\cite{lopez,kaser} or a cluster-based linear regression model
\cite{trivedi}. Special problems and suitable approaches for clustering 
structured educational data are surveyed in \cite{whedclust}.

Research on clustering educational texts and other non-struc\-tured data
is much more sparse, and we have been able to find only a few research
papers in which open responses from education-related questionnaires
were clustered. In \cite{yangmysore2009} a semi-supervised method
was adopted, where the answers were first clustered once, then a human
specialist identified the main topics from the preliminary clustering,
and finally, the answers were clustered again using the topics as
cluster representatives. In \cite{hirasawa2007} a fully-automatic
two-phase method was proposed, in which a preliminary probabilistic
clustering was first done with the EM-algorithm, and then the most accurately
clustered documents were used to determine parameters for the second-turn
EM-clustering. In addition, clustering has been utilized in grading 
text-formed exam answers \cite{basu,jing,wolska} and essays \cite{razon}.

Clearly, clustering open-ended questionnaire responses and other short
educational texts is an important but little researched problem. This
is not surprising, since clustering short texts is a difficult problem
in general, and new algorithms are not readily available. In this
paper we report an empirical comparison of three clustering methods,
$k$-means, affinity propagation and spectral clustering, on ten data
sets of students' open responses in English and Finnish. We compare
the results to human classifications and evaluate the effect of stemming
on clustering performance. In addition, we evaluate how well the main
topics of answers can be restored with cluster representatives.

The rest of the paper is organized as follows. In Section
\ref{secmethods} we survey the problems and methods of clustering 
short texts. In Section \ref{secexperiments} we describe
the materials and methods of our experiments. In Section
\ref{secresults} we present the results and discuss their meaning. The
final conclusions are drawn in Section \ref{secconcl}.

\section{Clustering short texts}
\label{secmethods}

Students' open-ended questionnaire responses are seldom clustered, but
we can expect that the methods for clustering other types of short
texts are applicable to them, too. In the following we survey the
problems and main approaches to short text clustering.

\subsection{Problems of short text clustering}

Short texts can be roughly divided into three categories by their
length: {\em word-level}, {\em sentence-level} and {\em
paragraph-level} documents. The word-level texts may contain just
one or a few words, like search engine queries and titles of search
results. The sentence-level texts contain one or more sentences, but
less than a paragraph. Well known examples of sentence-level documents include
micro-blogs, such as tweets and text 
snippets returned by search engines. Paragraph-level
documents contain usually just one paragraph, such as abstracts of
scientific papers. The length of questionnaire responses can be
anything from a word or two to a paragraph, but typically they contain
just one or two sentences.

The shortness of documents poses extra challenges to text
clustering. The main problem is that the traditional similarity
measures, like the well-known cosine similarity, rely heavily on the
term co-occurrence between documents. This means that the similarity
is not detected, if the documents do not share common terms, even if
they were topically related. The lack of common terms is most likely
in domains where the documents are short but vocabularies are large
\cite{pinto2007}, like tweets. However, the shortness itself does not
necessarily mean lack of co-occurring terms and there are successful
examples of using the cosine measure even for word-level texts
\cite{kang2010,niquan2011,yangmysore2009}. Course feedback answers
tend to have rather limited vocabularies and the same keywords often occur
in many answers. Therefore, it is expected that the similarity between
documents can be estimated from their word contents, after filtering 
irrelevant words.

Another commonly mentioned problem of short texts is the lack of
context \cite{metzler}. A long text usually offers a context for the
correct interpretation of the word, while short texts may share the
same words but still be topically unrelated. However, we recall that
the surrounding text is not the only context a word has. For example,
the questionnaire responses share the same context defined by the
question and questionnaire. In addition, the responders to educational
questionnaires usually have a common background: they may have
participated on the same course, read the same material, or tried to solve
the same problems. Therefore, the context can actually be very
specific, even if the answer contains just one word.

A third problem is that very short texts tend to be {\em noisy}, i.e.,
they often contain slang and other imprecise expressions, contracted
forms of words, and relatively many typographical errors
\cite{anastasiu2013}. This property is also common to questionnaire
answers \cite{yangmysore2009}.

A fourth problem is that short documents, like questionnaire
responses, often contain many outliers \cite{yangmysore2009}. This
property does not concern just textual data but educational (student)
data, in general, and should be taken into account in the selection
of clustering methods \cite{whedclust}.

\subsection{Approaches to short text clustering}

The standard approach for text clustering is the following: First, the
documents are represented in the {\em vector space model}, where each
document is considered to be a set of words and is represented as a
vector in the term space. The elements of document vector
$\mathbf{d}=(d_1,\hdots,d_m)$ can be simple boolean values (occurrence
of term $d_i$ in document $\mathbf{d}$), frequencies of terms $d_i$ or
their weighted transformations. The most popular approach is to
represent the vectors in the {\em tf-idf scheme}, where each element
$d_i$ is the term frequency (tf) weighted by its inverse document
frequency (idf). This scheme decreases the weight of frequent (and
poorly discriminating) terms. In addition, it is advisable to
normalize the vectors to unit length, to prevent the dominance of long
documents. This is especially important with short texts, where the
relative differences in document length can be substantial. 

In the preprocessing phase, the standard operations are stop word
filtering and stemming. In addition to stop words (lexicon specific
frequent terms), other overly frequent (uninformative) terms can be
removed, as well as very rare words. This reduces the data
dimension. In stemming, the word suffices are removed according to
certain heuristics, for deriving the word base. Alternatively, one can
transform the word into its basic form using dictionaries
(lemmatization). Stemming can reduce the data dimensionality
substantially, and it is generally believed to improve the clustering
quality. However, our experiments with questionnaire response data
show that the effect may sometimes be detrimental.

Usually, the data dimensionality is still too large for distinguishing
clusters (or close and far points). On the other hand, it is known
that the words are often highly correlated and most of them are
redundant for clustering. Therefore, it may be beneficial to reduce the
dimensionality further, either by feature extraction before clustering
(like principal component analysis or latent semantic analysis) or by
using clustering methods (like spectral clustering) which perform an
implicit dimension reduction.

Selection of a distance or similarity measure is considered
crucial for clustering ordinary data, but in the case of text
documents, the vocabulary and dimensionality seem to play a much
bigger role \cite{shrestha2012}. 
Selection of the clustering method is probably a more important issue,
but there are no comprehensive comparisons between different methods
for short texts. However, adjacency-based methods (like spectral
clustering) have generally worked well with text data. There is
anecdotal evidence that spectral clustering would be a good choice
(better than the tested hierarchical methods) for short texts as
well, as long as the clusters are not overly unbalanced
\cite{shrestha2012}. Affinity propagation is another method which has
produced promising results with short texts
\cite{kang2010,rangrej}. 

A common special technique used in short text clustering and
information retrieval is document expansion. The underlying idea is to
create the missing context by augmenting short texts, for example with web
search results. However, with questionnaire answers there is already
an implicit context and it is not clear whether expansion techniques
could produce any added value. It is equally possible that the
expansion could introduce only noise, especially when the
open responses concern personal opinions, attitudes, emotions, and
motivation. However, some external sources like taxonomies of
domain-specific terms \cite{liding2008} could well help in reducing
the dimensionality and categorizing the answers.

\section{Experiments}
\label{secexperiments}

The main objective of the experiments was to compare the most
promising clustering techniques for open responses. In addition, we 
evaluated how well the cluster representatives matched the main topics of 
answers.

\subsection{Data and preprocessing}

The data sets and their characteristics are described in Table
\ref{questiontable}. Each data set contained students' free-formed
answers to an open-ended question in a course feedback query. The
answers were collected from four different courses: Communication
Skills, IT skills, Law and Ethics ($Q1$--$Q3$), Introductory Security
($Q4$--$Q5$), Theoretical Foundations of Computer Science
($Q6$--$Q8$), and Programming ($Q9$--$Q10$). The first five sets of
answers ($Q1$--$Q5$) were in English, collected in the University of Warwick, 
and the last five sets ($Q6$--$Q10$) in Finnish, collected
in the University of Eastern Finland.

In the preprocessing phase the most frequent stop words were removed
and the words were stemmed. We refer to the three versions of the data
as the {\em raw data} (no processing), the {\em filtered data} (only the stop
words removed) and the {\em stemmed data} (stemmed version of the filtered
data). The stemming of the English data was done with the Malaga tool
\cite{malaga} and of the Finnish data with Voikko-fi (old
Suomi-malaga) \cite{voikkofi}.

Preprocessing decreased the vocabulary size, especially in
Finnish data sets and stop word removal decreased the average length
of answers (Table \ref{questiontable}). We note that in the Finnish
data sets, the vocabularies were much larger and answers were longer than
in the English data sets (especially questions $Q8$ and $Q9$).

\begin{table*}[!ht]
\centering
\caption{Used questions from the course feedback queries. Each data set is described by the number of answers ($\nA$), average length of answers and the number of unique words in the original data ($\avgA$, $\nW$) and in the preprocessed data after removing stop words and stemming ($\avgS$, $\nSW$).}
\label{questiontable}
\begin{tabular}{|ll|r|r|r|r|r|} 
\hline
&Question&$\nA$&$\avgA$&$\nW$&$\avgS$&$\nSW$\\
\hline
$Q1$&List topics you found particularly difficult and state why&46&9.7&214&6.7&159\\
$Q2$&List the best features of the module (up to 3), and&40&8.1&169&5.8&112\\
&indicate why&&&&&\\
$Q3$&How could we improve the module (up to 3 suggestions)?&42&11.1&232&9.4&202\\
$Q4$&List topics you found particularly difficult and state why&36&11.3&183&7.2&119\\
$Q5$&List the best features of the module (up to 3), and&32&11.8&161&7.8&125\\
&indicate why&&&&&\\
$Q6$&Why did you (not) go to the lectures?&39&14.6&365&11.8&228\\
$Q7$&Why did you (not) go to the exercise groups?&38&10.9&304&8.7&178\\
$Q8$&Feedback and improvement suggestions?&42&48.3&1089&40.4&658\\
$Q9$&Have you programmed before? What programming&95&27.2&1036&23.9&613\\
&languages you can use and how well?&&&&&\\
$Q10$&Have you had any problems during the course?&95&12.9&627&10.2&388\\
\hline
\end{tabular}
\end{table*}

\subsection{Human classification}

Before algorithmic clustering, reference classifications were created
by human experts. The classes were defined according to the main
themes that occurred in answers and one answer could belong to several
overlapping classes. In the evaluation, overlapping classes were
interpreted as a probabilistic classification: if an answer belonged
to $m$ classes, the probability of it belonging to any of these
classes was $1/m$. If an answer did not fit any natural class
(presented a unique theme), it was left as an outlier (a class of its
own).

The human classifications are described in Table \ref{humanclust}. The
number of proper classes (with $\geq 2$ answers) was relatively small
(3--6), but outliers were common (nearly 24\% of answers in
$Q3$). There was also considerable overlapping between classes and in
an extreme case ($Q2$), nearly 28\% of answers belonged to multiple
classes. Another extreme was $Q9$ that contained no outliers or
overlapping classes. In this question the answers were classified
into exclusive ordinal classes according to the amount of programming
experience (much--none). In reality, many answers would lie between
two consecutive classes, but it was impossible to interpret the answers in
such detail.

\begin{table*}[!ht]
\centering
\caption{Description of human classifications: number of classes with $\geq$ 2 answers ($K$), 
number of outliers ($n_{ol}$), number of answers belonging to multiple classes ($n_{mc}$), and description of the main topics (themes of classes containing at least 4 answers). The total number of classes is $K+n_{ol}$. Abbreviation PBL = problem-based learning.}
\label{humanclust}
\begin{tabular}{|l|r|rr|rr|l|} 
\hline
data&$K$&~~$n_{ol}$&(\%)&~~$n_{mc}$ &(\%)&Main topics\\
\hline
$Q1$&6&2 &(4.3\%)&7 &(15.2\%)&Scripting, essay/writing, term 2, law\\ 
$Q2$&6&7 &(17.5\%)&11 &(27.5\%)&Scripting, presentations, Linux, essay, seminar\\
$Q3$&5&10 &(23.8\%)&2 &(4.8\%)&More help, less work, scheduling, lectures\\
$Q4$&3&3 &(8.3\%)&5 &(13.9\%)&Encryption, virtual machines\\
$Q5$&3&4 &(12.5\%)&4 &(12.5\%)&Lab sessions, lectures, virtual machines\\
$Q6$&4&1 &(2.6\%)&2 &(5.1\%)&Participated PBL, learnt in lectures, schedule\\ 
&&&&&&problems, other reasons for not participating\\ 
$Q7$&3&2 &(5.3\%)&4 &(10.5\%)&For learning, getting points\\ 
$Q8$&3&4 &(9.5\%)&3 &(7.1\%)&PBL good, good teacher, why traditional style\\
$Q9$&5&0 &(0\%)&0 &(0\%)&Amount of programming experiences (much--none)\\
$Q10$&6&4 &(4.2\%)&12 &(12.6\%)&No problems, Java compiler, submission system,\\ 
&&&&&&WebCT, Jeliot, learning\\ 
\hline
\end{tabular}
\end{table*}

\subsection{Computational clustering}

The computational clustering was done with three clustering methods:
$k$-means, affinity propagation, and spectral clustering. The
$k$-means algorithm was selected as a reference method. Affinity
propagation and spectral clustering were selected because they have
shown promising results in previous research and their
implementations were readily available (unlike more exotic methods).

For clustering, the answers were represented in the vector space model
using the tf-idf weighting scheme. Cosine similarity was used as the
similarity measure. All clusterings were performed with the
Scikit-learn tool \cite{scikit}. Affinity propagation determines the
number of clusters itself, but for the $k$-means and spectral
clustering, we determined the optimal number by the `elbow method'
(identified the $k$-value corresponding to the `elbow' in the MSE
graph).

All three clustering methods were applied separately for both the
filtered data (only stop words removed) and stemmed data (stemmed
version of the filtered data). This resulted in six clusterings for
each of the ten data sets (i.e., 60 clustering).

\subsection{Evaluating the clustering performance}

In the evaluation all 60 computed clusterings were compared to the
corresponding human classifications. The goodness of computational
clustering was evaluated with two goodness measures, purity
(`accuracy' in \cite{yangmysore2009}) and the normalized mutual
information by Strehl and Ghosh \cite{strehlghosh}. Both measures have
widely been used in previous research for evaluating text clustering
results.

Purity of clustering $\Omega=\{\omega_1,\hdots,\omega_M\}$ given classification $C=\{c_1,\hdots,c_L\}$ is defined as
\begin{equation}
purity(\Omega,C)=\frac{1}{N}\sum_{\omega_i\in \Omega} \max_{c_j\in C}|\omega_i \cap c_j|,
\end{equation}
where $N$ is the size of the data set $D=\{d_1,\hdots,d_N\}$. 
Purity measures the extent to which clusters contain answers from a single
class. When questionnaire answers are clustered, high purity reflects that 
the clustering managed to catch the main message from all classes.  
However, purity does not take into account the number of clusters
that present the same class. In a pathological case a clustering of
singletons (single element classes) obtains purity=1. Therefore, one
should also take into account the number of clusters or use other
quality criteria.

In our case the human classifications were probabilistic and therefore we used 
a modification
\begin{equation}
purity'=\frac{1}{N}\sum_{\omega_i\in \Omega} \max_{c_j\in C}\left\{\sum_{d\in\omega_i} w(c_j|d)\right\},
\end{equation}
where $w(c|d)=1/m$, if $d\in c$ and $w(c|d)=0$ otherwise, when  
$d$ belongs to $m$ classes. We note that now $purity'<1$ whenever some answers belong to multiple classes.

Normalized mutual information between clustering $\Omega$ and classification 
$C$ is defined as 
\begin{equation}
NMI(\Omega,C)=\frac{I(\Omega,C)}{\sqrt{H(\Omega)H(C)}},
\end{equation}
where $I$ is mutual information:
$$I(\Omega, C)=\sum_{\omega_i\in \Omega}\sum_{c_j\in C}P(\omega_i,c_j)\log \frac{P(\omega_i,c_j)}{P(\omega_i)P(c_j)}$$
and $H$ is entropy:
$$H(\Omega)=\sum_{\omega_i\in \Omega}P(\omega_i)\log P(\omega_i) \textrm{ and } 
H(C)=\sum_{c_i\in C}P(c_i)\log P(c_i).$$
Probabilities are usually estimated by relative frequencies (maximum likelihood 
estimates) in the data $D=\{d_1,\hdots,d_N\}$, $|D|=N$.
Since the human classifications were probabilistic we used modified equations 
$$P(\omega,c)=\frac{1}{N}\sum_{d\in \omega}w(c|d),$$
$$P(\omega)=\frac{|\{d~|~d\in \omega\}|}{N} \textrm{ and }P(c)=\frac{1}{N}\sum_{d\in D}w(c|d).$$
Once again, $\NMI$ could not obtain its maximum value $\NMI=1$, since 
a hard clustering and a probabilistic classification could never be identical.

$\NMI$ is a popular validation measure in clustering studies since it
avoids drawbacks of many other measures, in particular it is independent
of the number of clusters and robust to small variations in
clusterings.  However, $\NMI$ has one well-known shortcoming: if the
true classification contains relatively many outliers (singleton classes)
or, alternatively, a `rag bag' class, where a human categorizer
would insert all such points, the $\NMI$-value becomes distorted
\cite{amigo2009,yangmysore2009}. Therefore, it can sometimes seriously
underrate the goodness of clusterings for open-form questionnaire
data.

\subsection{Extracting topics from the best clusterings}

Finally, we analysed how well the topics of the main clusters could be
extracted with the {\em HITS algorithm} ({\em Hyperlink-Induced Topic
  Search} \cite{hitsalg}).  HITS was applied to all main clusters
(containing at least four answers) to find the cluster representatives
(principal eigenvectors). In the HITS analysis, we used raw data,
since it had produced better results than filtered or stemmed data in
our earlier experiments.  In the evaluation, we analyzed how often the
cluster representative matched the major class of the cluster and
whether the representatives together covered all main topics of
answers, as shown in Table \ref{humanclust}.

\section{Results and discussion}
\label{secresults}

The results of comparisons between computational and human clusterings
are given in Table \ref{resulttable}.  
Overall, the clustering quality was quite good with both evaluation measures, 
taking into account small data sizes, relatively frequent
outliers, and overlapping clusters.
In addition, we recall that neither purity nor
$\NMI$ could obtain its maximum value due to probabilistic human
classifications.

Average purity of clusterings was 0.56--0.70 (whole range
0.37--0.82). For the average performance, there were no big differences
between English and Finnish data, but English data had higher purity
values, when the best clusterings for each question were considered.
Average $\NMI$ was 0.23--0.55 (whole range 0.13--0.66) and the values
were clearly larger for the English data (average 0.35--0.55, range
0.29--0.66) than the Finnish data (average 0.23--0.28, range
0.13--0.41). One possible reason for the difference between languages
is that the Finnish data contained larger vocabulary and the answers
were substantially longer and less focused.

For comparison, Yang et al. \cite{yangmysore2009} obtained average
purity of 64.4--70.9 and average $\NMI$ of 0.28--0.68 in similar but
much larger ($n=198-1196$) English data sets, when the answers were
clustered with three unsupervised clustering methods ($k$-means,
single-link hierarchical, and co-occurrence clustering). However, an
important difference to our case is that they used extremely large
cluster numbers ($k$=72--163) and both computational and human
clusterings contained many singletons. It was noted that this exaggerates both
the purity and $\NMI$ values. When singletons were excluded, the purity dropped to
16.5--44.0. In our clusterings, singletons were rare except in the
clustering of $Q10$ by affinity propagation. Consequently, the removal
of singletons changed the average purity values only little (average 0.57--0.65).

With both purity and $\NMI$ the most successful method in our
clusterings was affinity propagation (especially in English data sets)
but also $k$-means produced competitive results (especially in Finnish
data sets). Spectral clustering did not succeed particularly well,
except in $Q10$. One possible reason for the success of affinity
propagation is that it determines the optimal number of clusters
automatically. In our experiments this number (6--22) was always at
least as large as the number of clusters determined by the elbow
method (6--9) for $k$-means and spectral clustering. However,
sometimes this property can also be a weakness, as demonstrated by
$Q10$, where affinity propagation failed altogether after constructing
22 clusters, 20 of them singleton clusters. We note that the purity
was still relatively good (0.56) but decreased significantly (to 0.44)
when singletons were ignored.

\begin{table}[!h]
\centering
\caption{Results of the quality evaluation for the three clustering methods for the filtered and stemmed versions of data. The goodness measures are purity and $\NMI$. For each question, the best values have been emphasized.}
\label{resulttable}
\begin{tabular}{|l|r|r|r|r|r|r|r|r|r|r|r|r|} 
\hline
&\multicolumn{6}{|c|}{purity}&\multicolumn{6}{|c|}{$\NMI$}\\
&\multicolumn{2}{|l|}{$k$-means}&\multicolumn{2}{|l|}{aff. prop.}&\multicolumn{2}{|l|}{spectral}
&\multicolumn{2}{|l|}{$k$-means}&\multicolumn{2}{|l|}{aff. prop.}&\multicolumn{2}{|l|}{spectral}\\
&filt&stem&filt&stem&filt&stem&filt&stem&filt&stem&filt&stem\\
\hline
$Q1$&0.71&0.65&0.80&{\bf 0.82}&0.63&0.65
&0.53&0.39&0.60&{\bf 0.66}&0.34&0.38\\
$Q2$&0.52&0.53&0.46&{\bf 0.57}&0.47&0.51
&0.51&0.49&0.48&{\bf 0.58}&0.40&0.48\\
$Q3$&0.51&0.44&0.54&{\bf 0.56}&0.40&0.37
&0.52&0.47&{\bf 0.54}&{\bf 0.54}&0.35&0.34\\
$Q4$&0.74&0.74&0.79&{\bf 0.82}&0.71&0.74
&0.43&0.29&0.44&{\bf 0.48}&0.33&0.39\\
$Q5$&0.75&0.72&{\bf 0.81}&0.72&0.59&0.66
&0.51&0.47&{\bf 0.57}&0.51&0.32&0.39\\
\hline 
$\sum_{Eng}$&0.65&0.62&0.68&{\bf 0.70}&0.56&0.59
&0.50&0.42&0.53&{\bf 0.55}&0.35&0.40{\rule{0pt}{2.2ex}}{\rule[-1.2ex]{0pt}{0pt}}\\
\hline
$Q6$&0.64&{\bf 0.72}&0.62&0.62&0.56&0.67
&0.34&0.40&0.26&0.32&0.24&{\bf 0.41}\\
$Q7$&0.63&0.68&{\bf 0.71}&0.68&0.55&0.63
&0.28&0.20&{\bf 0.31}&0.21&0.19&0.19\\
$Q8$&0.74&0.73&0.73&{\bf 0.75}&0.70&0.70
&{\bf 0.34}&0.30&0.32&0.32&0.23&0.24\\
$Q9$&0.51&0.48&{\bf 0.62}&0.53&0.46&0.47
&0.22&0.17&{\bf 0.34}&0.26&0.20&0.23\\
$Q10$&0.59&0.60&0.56&0.56&0.62&{\bf 0.63}
&0.22&0.25&0.13&0.13&0.30&{\bf 0.31}\\
\hline
$\sum_{Fin}$&0.62&{\bf 0.64}&{\bf 0.64}&0.63&0.58&0.62
&{\bf 0.28}&0.26&0.27&0.25&0.23&{\bf 0.28}{\rule{0pt}{2.2ex}}{\rule[-1.2ex]{0pt}{0pt}}\\
\hline
$\sum_{All}$&0.63&0.63&{\bf 0.66}&{\bf 0.66}&0.57&0.60
&0.39&0.34&{\bf 0.40}&{\bf 0.40}&0.29&0.34{\rule{0pt}{2.2ex}}{\rule[-1.2ex]{0pt}{0pt}}\\
\hline
\end{tabular}
\end{table}

Maybe the most surprising result was how often stemming had a negative
or no effect on the clustering performance (13/30 cases with purity
and 16/30 cases with $\NMI$). This is an important discovery, since
stemming is a central part of text clustering and it is generally
believed to improve the performance. However, our results suggest that
stemming should be used with caution since it can deteriorate the
quality quite remarkably especially when measured with $\NMI$
(decrease of up to 32\%). This phenomenon affected all methods, but
spectral clustering seemed to benefit from stemming most consistently.

The HITS analysis was also more successful with the English data
sets. For the English data, the cluster representatives covered all the main
answer topics, as shown in Table \ref{humanclust}, except one small and
diverse topic in $Q3$ (on average 95\% of topics). This means that one
could restore the main answer topics by reading merely the
representatives of the main clusters. The cluster representatives also matched
the major classes of clusters in all except one small heterogenous
cluster in $Q2$. This suggests that HITS representatives summarized the
main topics of clusters well. For the Finnish data, the cluster
representatives covered all the main topics only in $Q6$ and $Q7$ (on
averge 76\% of main topics). In addition, the cluster representatives
matched the main topics only in 56\% of the main clusters.

\section{Conclusions}
\label{secconcl}

In this paper, we have presented our experiments on clustering ten
data sets of open responses from course feedback queries with three
clustering methods ($k$-means, affinity propagation, and spectral
clustering).  The results suggest that the combination of clustering
and topic extraction with the HITS algorithm can summarize the main
messages of students' feedback accurately at least in English
questionnairies. On average, the best clustering performance was
achieved with affinity propagation, but also $k$-means produced
competitive results.

The results showed a clear discrepancy between the English and Finnish
data sets. For the English data affinity propagation performed
well in all data sets despite frequent outliers and considerable
overlapping between real clusters. On the other hand, for the Finnish
data sets the performance was poorer and none of the methods clearly
outperformed others. The HITS analysis produced similar results. For
the English data, one could restore virtually all main topics of
answers by reading merely presentatives of the main clusters, but for
the Finnish data, nearly one quarter of topics were missed. It is possible that
the difference is partially explained by the larger vocabulary, longer
answers, and less focused questions in the Finnish data.

The most interesting discovery was that stemming often deteriorated
the clustering quality, sometimes dramatically. In future research, we
intend to study reasons for this behaviour and also experiment with
document expansion techniques that enrich the answers before
clustering.

\bibliographystyle{abbrv}
%


\end{document}